\documentclass[journal=jctcce,manuscript=article,layout=onecolumn]{achemso}

\usepackage{bm}
\usepackage{siunitx}
\usepackage[normalem]{ulem} 
\usepackage[sc,osf]{mathpazo}

\author{Valerio Rizzi}
\affiliation{Department of Chemistry and Applied Biosciences, ETH Zurich, c/o USI Campus, Via Giuseppe Buffi 13, CH-6900, Lugano, Ticino, Switzerland}
\alsoaffiliation{Facolt{\`a} di Informatica, Istituto di Scienze Computazionali, Universit{\`a} della Svizzera italiana (USI), Via Giuseppe Buffi 13, CH-6900, Lugano, Ticino, Switzerland}
\altaffiliation{Contributed equally to this work}

\author{Dan Mendels}
\affiliation{Department of Chemistry and Applied Biosciences, ETH Zurich, c/o USI Campus, Via Giuseppe Buffi 13, CH-6900, Lugano, Ticino, Switzerland}
\alsoaffiliation{Facolt{\`a} di Informatica, Istituto di Scienze Computazionali, Universit{\`a} della Svizzera italiana (USI), Via Giuseppe Buffi 13, CH-6900, Lugano, Ticino, Switzerland}
\alsoaffiliation{Institute for Molecular Engineering, University of Chicago, Chicago, Illinois 60637, United States}
\altaffiliation{Contributed equally to this work}

\author{Emilia Sicilia}
\affiliation{Dipartimento di Chimica e Tecnologie Chimiche, Universit\`a della Calabria, 87036 Rende (CS), Italy}

\author{Michele Parrinello}
\email{parrinello@phys.chem.ethz.ch}
\affiliation{Department of Chemistry and Applied Biosciences, ETH Zurich, c/o USI Campus, Via Giuseppe Buffi 13, CH-6900, Lugano, Ticino, Switzerland}
\alsoaffiliation{Facolt{\`a} di Informatica, Istituto di Scienze Computazionali, Universit{\`a} della Svizzera italiana (USI), Via Giuseppe Buffi 13, CH-6900, Lugano, Ticino, Switzerland}
\alsoaffiliation{Istituto Italiano di Tecnologia (IIT), Via Morego, 30, 16163 Genova GE, Italy}

\title[\texttt{achemso} Blind search for complex chemical pathways]
{Blind search for complex chemical pathways using Harmonic Linear Discriminant Analysis}

\begin{document}

\maketitle
\date{\today}
\begin{abstract}
\noindent
Disentangling the mechanistic details of a chemical reaction pathway is a hard problem that often requires a considerable amount of chemical intuition and a component of luck. Experiments struggle in observing short-life metastable intermediates, while computer simulations often rely upon a good initial guess. In this work, we propose a method that, from the simulations of a reactant and a product state, searches for reaction mechanisms connecting the two by exploring the configuration space through metadynamics, a well known enhanced molecular dynamics method. The key quantity underlying this search is based on the use of an approach called Harmonic Linear Discriminant Analysis which allows a systematic construction of collective variables. 

Given the reactant and product states, we choose a set of descriptors capable of discriminating between the two states. In order not to prejudge the results, generic descriptors are introduced. The fluctuations of the descriptors in the two states are used to construct collective variables. We use metadynamics in an exploratory mode to discover the intermediates and the transition states that lead from reactant to product. The search is at first conducted at a low theory level. The calculation is then refined and the energy of the intermediates and transition states discovered during metadynamics is computed again using a higher level of theory.

The method's aim is to offer a simple reaction mechanism search procedure that helps in saving time and is able to find unexpected mechanisms that defy well established chemical paradigms. We apply it to two reactions, showing that a high level of complexity can be hidden even in seemingly trivial and small systems. The method can be applied to larger systems, such as reactions in solution or catalysis.

\end{abstract}

\newpage

\section*{Introduction}

Reactions lie at the heart of Chemistry. Unveiling their mechanisms is of paramount importance for a vast number of applications such as increasing rates, yields, selectivity and, ultimately, for optimizing reaction conditions at a low energy cost and, hopefully, without jeopardizing the environment. 
Although experimental techniques can shed light on a large variety of aspects pertaining to chemical reactions, information about the mechanisms coming from experiments is limited because structures and energies of short-life intermediates and transition states continue to be experimentally elusive. 
In contrast, computational methodologies do not suffer from such limitations and allow the investigation of events at the atomic level and at the femtosecond timescale. Nevertheless, their use in the search for reaction pathways still needs human input in the form of an extensive chemical knowledge and intuition.

The standard procedures for mapping a chemical reaction are known to be hard and tedious, as a mix of chemical intuition, patience and a component of luck is required for properly taking into account and describing all the elementary steps composing the whole mechanism. The large dimensionality of the reaction space quickly makes tracking the multiple atomic movements a daunting task, even for very small systems. 
Over the years, numerous automated computational procedures have been developed with the ambitious aim of helping to elucidate reaction mechanisms in several and diverse systems \cite{Maeda2011,Maeda2013,Pietrucci2015a,Pietrucci2018,PerezDeAlbaOrtiz2018,Bergeler2015,Simm2019,Grimme2019}. A large subset of these methods finds connecting pathways between reactants and products by carrying out a search along the Minimum Energy Path (MEP) \cite{Weinan2002,Weinan2005,Henkelman2000,Peters2004,MallikarjunSharada2012}, but these methods suffer from the inability to take into account entropic contributions. In addition, their effectiveness \cite{Weinan2002,Weinan2005,Henkelman2000,Peters2004,MallikarjunSharada2012} often relies on an initial guess for the reaction pathway \cite{Borrero2016} or on the preliminary knowledge of intermediate states. When such input is not available, these methods might tend to locate the geometrically shortest pathways between reactants and products, rather than those characterized by lower barriers for the transition under examination \cite{Maeda2013}. 

Methods that perform their search on the system's Free Energy Surface (FES), thus including entropic contributions, have also been introduced \cite{Weinan2002,Weinan2005,Maeda2011,MallikarjunSharada2012,Wang2016,Shiga2018}. Such methods exploit \textit{ab-initio} molecular dynamics (MD) simulations directly, for example by running simulations at elevated temperatures or adding artificial forces to induce reaction activity \cite{Wang2016}. However, while such methods are able to discover reactions in a given system and \textit{a posteriori} map their corresponding mechanisms, they may alter the system FES of the process \cite{Wang2016}, be limited to unimolecular \cite{Muller2002} or bimolecular reactions \cite{Maeda2011} or are unable to focus on a single specific reaction \cite{Dewar1984,Pietrucci2011,Wang2016,Fu2018,Shiga2018}.

In this paper we propose a new approach that helps to speed up the discovery of complex reaction mechanisms, using a minimal amount of information. 
The method only requires MD simulations of the reactant (R) and product (P) states. Based on this information, a search is carried out along the direction that maximally separates R and P.
This direction is determined through a recently developed methodology, Harmonic Linear Discriminant Analysis (HLDA), that only uses the thermally induced fluctuations in the reactant and the product state.

States R and P are identified by a set of descriptors whose fluctuations are used to define the HLDA collective variable (CV). In the spirit of the present work that aims at obtaining an automatic procedure, we define a set of general descriptors, especially suited at monitoring the bond breaking and bond forming processes that take place in chemical reactions. The physical picture underpinning HLDA is the classical one of the rare event scenario in which there are two states separated by a high barrier and the system undergoes transitions between them. The HLDA variable is optimized to distinguish the two states and can be used in conjunction with enhanced sampling methods, such as Metadynamics (MetaD), to accelerate the transition rate between states. 

While in other papers \cite{Mendels2018,Piccini2018a,Mendels2018a,Rizzi2019,Zhang2019} the HLDA CV has been shown to succeed in mapping the FES of a variety of chemical reactions, in this paper we want to show that the HLDA procedure can be applied in a more general and exploratory mode. Here, in order to go from state R to P, the system has to cross a number of possibly unknown intermediate states. If the FES needs to be reconstructed, a common strategy is to use MetaD in conjunction with a gentle bias deposition to encourage back and forth transitions between basins. In contrast, for the aim of this work, we use a more aggressive schedule in order to achieve a rapid exploration of configuration space. The choice of an intense bias deposition has the consequence to promote forward reactions along the HLDA variable, while hindering backward reactions, forcing the system to find a path connecting state R and P. 

\begin{figure}[tb]
	\centering
	\includegraphics[width=0.85\columnwidth]{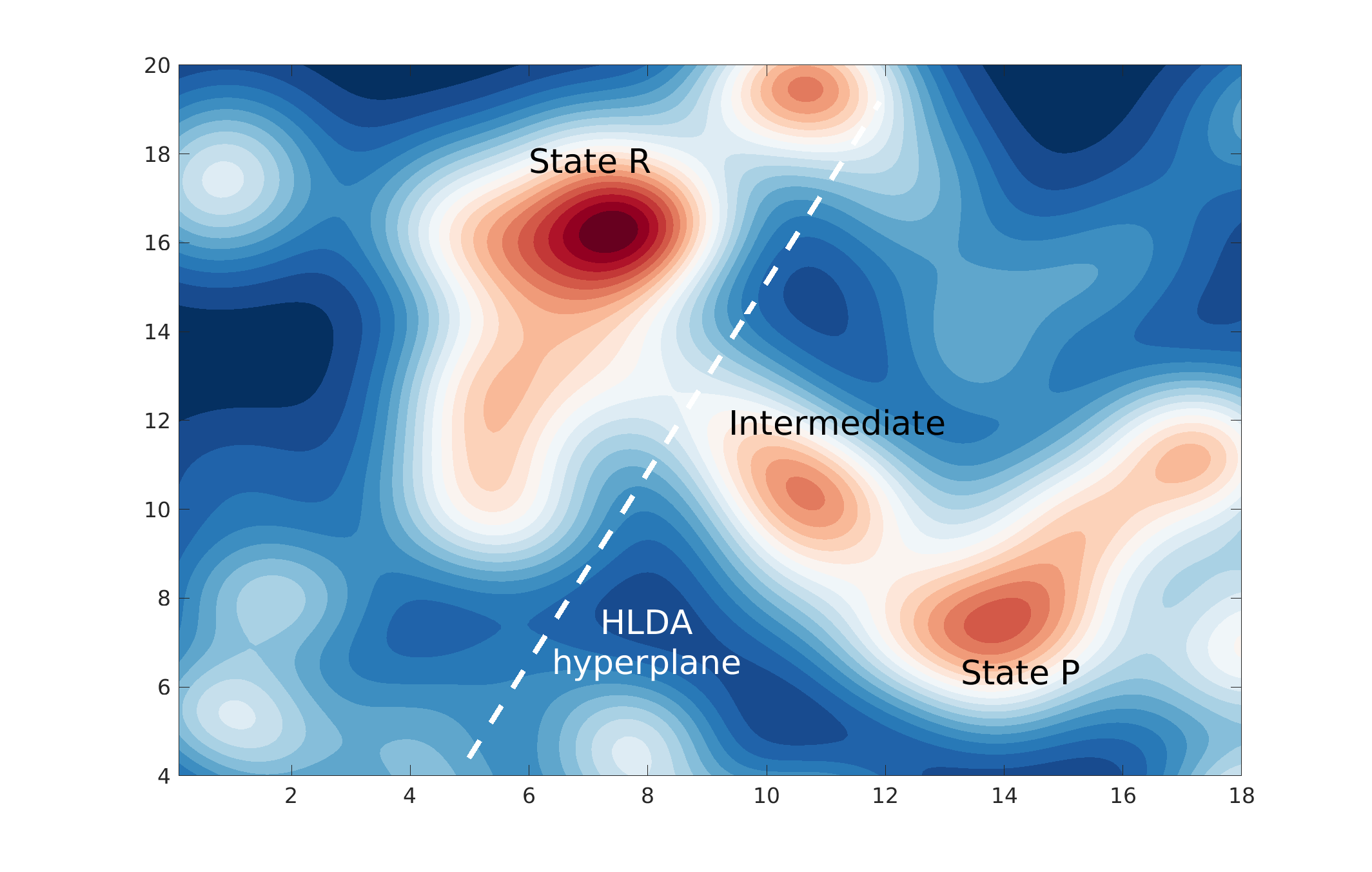}   
	\caption{A two dimensional illustration of the rational underlying the proposed HLDA method. Using this method, the system is encouraged to move from the reactant state R in a perpendicular direction to the HLDA hyper-plane, depicted in dashed line. Given the very uneven landscape which characterizes a chemical system reaction network, the system, rather than explore its own phase space indiscriminately, is encouraged to follow low free energy pathways leading, potentially through metastable intermediate states, to the product state P. 
	}
	\label{fig:Illustration_FE}
\end{figure}

In order to get an intuitive grasp of this reaction exploration procedure, we can take a look at Fig. \ref{fig:Illustration_FE} where, on an hypothetical FES, the pathway from state R to P has to go through an intermediate metastable state. In the same picture, we show one HLDA derived surface in which the CV has a constant value. By construction, the HLDA variable ensures that state R and P are optimally separated when projected into the direction orthogonal to the constant CV planes. Chemical reaction pathways occupy a very small fraction of the system FES landscape, especially if the FES is embedded in a high dimensional space of descriptors. In this landscape, the energy barriers surrounding state R tend to be very high as the process of bond breaking and forming is costly. It is reasonable to assume that in the general direction indicated by HLDA lies a chemical reaction pathway whose barrier is lower.

Therefore, depositing bias in the HLDA CV direction favors an exploration of a number of paths which is analogous to a sweep of configurational space in the region between the two states. Eventually, after a long enough exploration time, this sweep pushes the system across a barrier. This transition is analogous to entering a trap: the system is encouraged to proceed forward along the HLDA direction as it is surrounded by high barriers and is discouraged to go back because of the previous bias deposition.
The trap is akin to a funnel in configuration space as, while the system approaches the transition state, the number of available configurations at a given temperature is reduced.

In the simplest case, the reaction from state R to P occurs and the real-time trajectory of the transition indicates a possible pathway. On the other hand, if a transition to an intermediate occurs, two scenarios are possible depending on whether or not the HLDA variable is able to discriminate the newly found state. In the first case, the simulation can proceed forward, continue its exploration along the HLDA variable and eventually progress further. In the second case, it is recommendable to stop the simulation and include information about the intermediate state in the next set of simulations, either by iterating the HLDA procedure between the newly found intermediate and state P or by running multiclass HLDA \cite{Piccini2018a}. In the examples studied below we explore both possibilities. 

The level of electronic accuracy of the calculations is deliberately kept at a low level out of computational expediency. Besides, given the fact that the MetaD procedure has been accelerated, it is not possible to obtain converged FES from the HLDA simulations.
Instead, we can extract from the trajectories intermediate and transition state geometries. This information is fed into a standard quantum-chemical code that allows to raise the level of electronic structure theory to the required level. Such static simulations help in quantitatively mapping the reaction energy profiles and obtain an optimized estimate of the corresponding intermediate and transition states structure.

The input from HLDA attenuates one of the main stumbling blocks of standard approaches: the need for a guess of the reaction pathway. This limitation manifests itself even for some of the smallest systems and its resolution often requires a significant amount of human time and chemical intuition. The HLDA discovery helps in reducing the time consuming aspect, while requiring close to no chemical intuition. Furthermore, it is able to observe unexpected paths that defy conventional chemical paradigms, as it does not need rely upon any guess.

In order to test the method, we applied it to two multi-molecular reactions in gas phase and we found that in both cases it was able to unveil the reaction mechanisms rationalizing all the information reported in the literature as shown by the sequence of intermediates.
Static quantum-mechanical simulations carried out starting from the HLDA result confirmed that the reaction path goes through the intercepted intermediates and provided a better estimation of the corresponding energy barriers.

\section*{Methods}

MetaD \cite{Laio2002,Barducci2008,Valsson2016} is one of the key elements in our reaction discovery procedure as it allows accelerating the sampling of rare events in molecular dynamics simulations. The more advanced Well-Tempered version of MetaD \cite{Dama2014} improves the capability of mapping Free Energy Surfaces, but is not needed for our exploratory purposes. Here we use the original version as discussed in \cite{Laio2002,Barducci2008} and, for completeness, we recall its main features.

In this version, a bias potential $V(\bm{s})$ that is a function of a set of $N_s$ collective variables (CVs) $\bm{s}(\bm{R})$ is introduced to steer the system away from local minima and cross energy barriers. The bias potential is built iteratively, usually by depositing Gaussian kernels at fixed time intervals during the dynamics. The form of such kernel at the n-th iteration is
\begin{equation}
G(\bm{s},\bm{s}^{n}) = W \exp \left( - \sum_{k=1}^{N_s} \frac{(s_k - s_{k}^{n} )^2}{2 \, \sigma^2} \right)
\label{eq:kernel}
\end{equation}
where $W$ indicates the height of the Gaussian, $\sigma$ its standard deviation and $\bm{s}_{k}^{n}$ represents the value of the k-th CV at the n-th iteration.

A limitation of MetaD is that it does not scale well with an increasing number of CVs (usually not more than two), so the choice of few optimal CVs is crucial for the ability to investigate rare events such as chemical reactions. This choice is very time consuming and often requires a profound knowledge on the reaction at hand. Moreover, when the unknown reaction involves the concerted movement of several atoms, guessing an effective CV quickly becomes an insormountable task as the system size grows.

The aim of this paper is to employ the recently developed HLDA method \cite{Mendels2018} and devise a fairly general procedure that allows a reaction pathway to be reconstructed with a minimum of human effort and without prior knowledge.
The HLDA method acts as a dimensionality reduction tool, by determining from a set of descriptors $\bm{d}(\bm{R})$ a single CV $s_{H}(\bm{R})$ that optimally separates free energy minima. That variable can be used as a CV in MetaD, simplifying the kernel in Eq. \ref{eq:kernel} into $G(s_{H},s_{H}^{n}) = W \exp \left( \frac{(s_H - s_{H}^{n} )^2}{2 \, \sigma^2} \right)$.

The method's success still depends very much on the suitability of the descriptors, but it does not present any significant limitation on the number of descriptors $N_d$. This improvement allows the choice of a rather general descriptor set, that is not reaction-specific anymore and can be used in a number of different reactions. Before introducing our proposal for such a set, we provide a brief overview of the HLDA method itself.

Given a set of descriptors $\bm{d}(\bm{R})$, the HLDA procedure requires their statistical distribution from unbiased simulations in the R and P states. 
The expectation values $\boldsymbol{\mu}_{R}$ and $\boldsymbol{\mu}_{P}$ of the descriptors and the multivariate covariance matrices $\boldsymbol{\Sigma}_{R}$ and $\boldsymbol{\Sigma}_{P}$ between every couple of descriptors are the input for the HLDA procedure.

From these elements, HLDA builds a CV that corresponds to the direction $\mathbf{W}$ in the $N_d$-dimensional descriptor space that best separates state $R$ and $P$ \cite{Mendels2018}.
The direction $\mathbf{W}$ is computed by maximizing the ratio between the system so-called between class $\mathbf{S}_b$ and within class $\mathbf{S}_w$ scatter matrices. The former is estimated by the square of the distances  between the projected means, and can be written as $\mathbf{W}^T \mathbf{S}_b \mathbf{W}$
with
\begin{equation}
\label{between_class}
\mathbf{S}_b = \left( \boldsymbol{\mu}_R - \boldsymbol{\mu}_P \right)\left( \boldsymbol{\mu}_R - \boldsymbol{\mu}_P \right)^T.
\end{equation}
The latter is estimated from the harmonic average of the two states covariance matrices $\boldsymbol{\Sigma}_{R(P)}$, $\mathbf{W}^T \mathbf{S}_w \mathbf{W}$
where 
\begin{equation}
\label{harmonic_mean}
\mathbf{S}_w = \frac{1}{\frac{1}{\boldsymbol{\Sigma}_R} + \frac{1}{\boldsymbol{\Sigma}_P}}.
\end{equation}

\noindent
The HLDA objective function, which has the form of a Rayleigh ratio
\begin{equation}
\label{fisher_ration}
\mathcal{J(\mathbf{W})} = \frac{\mathbf{W}^T \mathbf{S}_b \mathbf{W}}{\mathbf{W}^T \mathbf{S}_w \mathbf{W}}
\end{equation}
is then maximized by
\begin{equation}
\label{maximizer}
\mathbf{W}^* = \mathbf{S}_w^{-1} \left( \boldsymbol{\mu}_R - \boldsymbol{\mu}_P \right).
\end{equation}
which, at last, yields the HLDA CV
\begin{equation}
\label{maximizer_harm}
s_{H}(\mathbf{R}) = \left( \boldsymbol{\mu}_R - \boldsymbol{\mu}_P \right)^T \left( \frac{1}{\boldsymbol{\Sigma}_R} + \frac{1}{\boldsymbol{\Sigma}_P} \right)   \bm{d}(\mathbf{R}).
\end{equation}
With the help of expression (\ref{maximizer_harm}), for any configuration $\bm{{R}}$ and its descriptor values $\bm{d}(\mathbf{{R}})$, one is able to calculate the corresponding HLDA variable $s_{H}(\bm{{R}})$. The last element missing is the selection of a suitable descriptor set $\bm{d}(\mathbf{{R}})$.

In principle, any quantity that captures the fluctuations of state R and P and is able to separate them could be chosen as a descriptor. 
However, in the present context, it is preferable that the set of descriptors is determined through a general strategy that does not prejudge the outcome of the simulation.
The set that we propose in this work can be used quite generally for a large variety of chemical reactions and could eventually be further generalized and extended. 

It is based on the concept of coordination number (CN), a quantity that takes the distances between groups of atoms and evaluates the presence of bonds through a switching function with a finite width. We use the expression
\begin{equation}
\label{Eq:Coordination_number}
q_{ij}(r_{ij})=\frac{1-\big(\frac{r_{ij}}{r_0}\big)^n}{1-\big(\frac{r_{ij}}{r_0}\big)^m}
\end{equation}
where $r_{ij}$ measures the distance between atoms $i$ and $j$, the parameter $r_0$ is calibrated in accordance with a typical bond length (if a bond is present) and $n$ and $m$ determine the steepness of the switching function. It has been observed that a rather soft switching function \cite{Bonomi2008} that presents a non-zero signal even at large distances is a good choice. Following this prescription, for our examples we have chosen $n=6$ and $m=8$. 

In a chemical reaction, bonds can be broken or formed, and, as a consequence, the CN between the atoms involved can increase or decrease accordingly. 
Out of the atom-specific $q_{ij}$, we want to construct descriptors that are permutationally invariant so as not to prejudge the reaction mechanism by a targeted variable.
For this reason, in a system that is composed of $N_S$ species, we introduce the following set of $2 N_S^2$ descriptors. For every couple of atomic species $A-B$, we compute the minimum and the maximum CN
\begin{equation}
\label{Eq:Min_coordination}
d_{\mathrm{min}}^{A-B}(\bm{R})=\lambda_{\mathrm{min}} \left(\log{\sum_{i \in A}  \exp\left(\frac{\lambda_{\mathrm{min}}}{  \sum_{j \in B} q_{ij}  }\right)  }\right)^{-1}
\end{equation}
\begin{equation}
\label{Eq:Max_coordination}
d_{\mathrm{max}}^{A-B}(\bm{R})=\lambda_{\mathrm{max}} \ \log{\sum_{i \in A} \exp\left(\frac{  \sum_{j \in B} q_{ij} }{\lambda_{\mathrm{max}}}\right)}
\end{equation}
where $\lambda_{\mathrm{min}}=20$ and $\lambda_{\mathrm{max}}=0.02$ are constants that regulate the sharpness of the functions.

It is natural to assume that bond forming processes tend to occur when the distance $ r_{ij}$ between two unbound atoms ${i \in A}$ and ${j \in B}$ gets shorter and $q_{ij}(r_{ij})$ gets larger, so the dynamics of $d_{\mathrm{max}}^{A-B}$ would be able to capture such an event. Analogously, $d_{\mathrm{min}}^{A-B}$ would be suitable to describe bond breaking processes. 
Our strategy consists of taking all couples of atomic species and use expressions (\ref{Eq:Min_coordination}) and (\ref{Eq:Max_coordination}) as our general set of descriptors $\bm{d}(\bm{R})$. We point out that this set also includes couples of the same species such as $d_{\mathrm{max}(\mathrm{min})}^{A-A}$ and all permutations, as in principle $d_{\mathrm{max}(\mathrm{min})}^{A-B} \ne d_{\mathrm{max}(\mathrm{min})}^{B-A}$.

If $N_A$ is the number of atoms in the systems belonging to species $A$ and $N_B$ the one corresponding to species $B$, this procedure is valid in the case of $N_A > 2$ and $N_B > 2$. If $N_A = 2$, it would make no sense taking both the maximum and minimum in $d_{\mathrm{max}(\mathrm{min})}^{A-A}$ as they would be identical, so only one of them is used as a descriptor. Furthermore, if $N_A = 1$, descriptors of the form $A-A$ would be meaningless, while $d_{\mathrm{max}}^{A-B} \approx d_{\mathrm{min}}^{A-B}$, so only one of the two is included. In some cases, excluding the CN of Hydrogen with itself is preferable given that Hydrogen molecule bonds are very short and tend to play an overly dominant role in the search dynamics.

The presence of a fairly large set of descriptors, is beneficial to the reaction discovery process as it allows to conduct the search in a multi-dimensional space. The most important descriptors for a reaction would present large contributions in the expression (\ref{maximizer_harm}), while descriptors with little relevance would provide near-zero contribution.

It is good practice to normalize every descriptor $d_k$ by its maximum value in both states during the unbiased dynamics $D_k = \max_{R,P} \left(d_k(\mathbf{R})\right)$, so that the amplitude of the respective fluctuations is comparable.
The standardized form of each descriptor $k=1,\dots,N_d$ can then be written as
\begin{equation}
\label{maxD}
\hat{d}_k(\mathbf{R})_{R(P)} = \frac{d_k(\mathbf{R})_{R(P)}}{D_k}.
\end{equation}

A problem that can arise is that a large set of descriptors may present a degree of correlation which can interfere with the HLDA procedure. 
Usually, a sign that correlation is interfering with the HLDA procedure is that most of the components of vector $\left( \boldsymbol{\mu}_R - \boldsymbol{\mu}_P \right)^T \left( \frac{1}{\boldsymbol{\Sigma}_R} + \frac{1}{\boldsymbol{\Sigma}_P} \right)$ from Eq. \ref{maximizer_harm} are close to zero.
To ensure that correlations are correctly filtered out, the application of a filtering protocol such as the singular-value decomposition (SVD) is recommended. This screening is applied to the two covariance matrices $\boldsymbol{\Sigma}_{R(P)}$.
According to SVD, these matrices can be written as
\begin{equation}
\label{svd}
\boldsymbol{\Sigma}_{R(P)} = \boldsymbol{U}_{R(P)} \boldsymbol{S}_{R(P)} \boldsymbol{V}_{R(P)}^*
\end{equation}
where $\boldsymbol{U}_{R(P)}$ and $\boldsymbol{V}_{R(P)}$ are unitary $N_d \times N_d$ matrices, while $\boldsymbol{S}_{R(P)}$ is a diagonal non-negative real-valued $N_d \times N_d$ matrix. Its diagonal values are known as singular values and they measure correlations, with the smallest ones corresponding to the most correlated descriptors. By setting the smallest $N_f$ singular values to zero, we build the new diagonal matrices $\boldsymbol{\tilde{S}}_{R(P)}$ and, from them, we obtain the filtered covariance matrices
\begin{equation}
\label{svdfiltered}
\boldsymbol{\tilde{\Sigma}}_{R(P)} = \boldsymbol{U}_{R(P)} \boldsymbol{\tilde{S}}_{R(P)} \boldsymbol{V}_{R(P)}^*.
\end{equation}
By inserting them in Eq. \ref{maximizer_harm}, we obtain a filtered HLDA CV
\begin{equation}
\label{maximizer_harm2}
\tilde{s}_{H}(\mathbf{R}) = \left( \boldsymbol{\mu}_R - \boldsymbol{\mu}_P \right)^T \left( \frac{1}{\boldsymbol{\tilde{\Sigma}}_R} + \frac{1}{\boldsymbol{\tilde{\Sigma}}_P} \right)   \bm{\hat{d}}(\mathbf{R}).
\end{equation}

We can start exploring the reaction pathway by applying a bias to $\tilde{s}_{H}$, running MetaD simulations from the reactant state. In Fig. \ref{fig:dynHLDA} in the Results section, we will show the dynamics of the HLDA variable during an unbiased and a MetaD simulation.
To encourage the exploration away from state R in the direction of P, we tend to choose a frequent bias deposition with a high $W$ in Eq. \ref{eq:kernel}. Simulations are run until a sharp change of $\tilde{s}_{H}$ is observed, indicating a reactive activity. If, upon inspection, the newly found state does not correspond to the expected reaction product state, we indicate two possible routes.

If the HLDA variable value of the newly found state does not overlap with the final reaction product, the MetaD simulation can proceed forward on its route towards discovery. On the other hand, if there is a significant overlap, the procedure can be iterated. In that case, the newly discovered state is utilized as a new effective reactant and a new HLDA variable is constructed linking it to the product state. The procedure can be repeated, if needed, for successive metastable states, until the system reaches the final product state. Another possibility is to employ multiclass HLDA \cite{Piccini2018a}, a procedure in which the newly found state is used together with the reactant and product states to determine two maximal separation HLDA directions. These directions can then be used as CVs in MetaD. The concomitant bias of multiple CVs makes the computational cost grow significantly, so such approach is limited to a only a few CVs, typically two.

The trajectory obtained during the MetaD simulation can be employed to extract a guess for the transition states. One can choose a set of configurations located within the sharp HLDA variable transitions and perform a committor analysis on them. This analysis consists of running short unbiased simulations starting from each of these configurations and counting the number of transitions forward and backward. The configurations that present a roughly equal number of these transitions represent a good estimate for the geometry of the corresponding transition state.

\begin{figure}[tb]
	\centering
	\includegraphics[width=1.0\columnwidth]{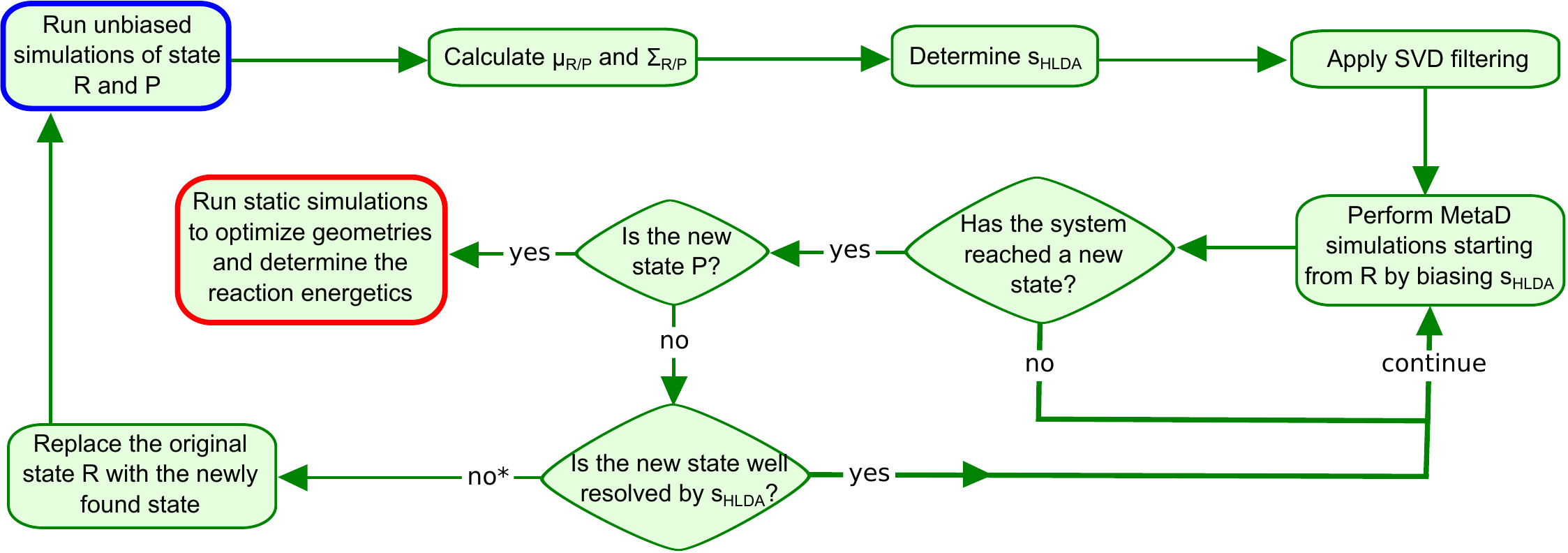}   
	\caption{Protocol of a typical HLDA reaction search between state R and P. The starting point is framed in blue, while the desired end point is framed in red. At the step with *, the multiclass HLDA can also alternatively be used (see SI for details).
	}
	\label{fig:protocol}
\end{figure}

At last, our guess for the intermediate and the transition states can be fed to a higher accuracy electronic structure calculation to obtain the reaction energetic profile and the optimized geometries corresponding to the full reaction pathway. The accuracy of such calculation can be tuned to the desired precision. So, in summary, the whole reaction pathway search procedure typically follows the protocol in Fig. \ref{fig:protocol}.

\section*{Computational details}
	 	 	 	 	
We performed the HLDA simulations with the \textit{ab-initio} molecular dynamics code CP2K, \cite{VandeVondele2005} using the PBE functional \cite{Perdew1996} and a DZVP basis set. The temperature was kept constant at 400 K through a velocity rescaling thermostat with a time constant of 50 fs \cite{Bussi2007}. 
The typical length for the preliminary unbiased runs was about 100 ps. The PLUMED plugin \cite{Tribello2014} was used for performing MetaD with the HLDA CV. Further details about such simulations can be found in the Supporting Information.

The static quantum-mechanical calculations were performed employing the Gaussian09 software package \cite{g09} in the framework of the density functional theory. The hybrid three-parameters B3LYP \cite{Lee1988,Becke1993} functional was used. For the description of $\mathrm{I}$ atoms, the relativistic compact Stuttgart/Dresden effective core potential \cite{Andrae1990} was adopted together with its split valence basis set. The standard 6-311G** basis set of Pople was employed for the rest of the atoms. Vibrational analysis for each optimized stationary point was carried out to determine its minimum or saddle point nature and to calculate zero-point vibrational energy corrections. For all intercepted transition states the vibrational mode associated with the imaginary frequency was carefully checked to correspond to the correct movement of the involved atoms. Furthermore, the IR \cite{Fukui1970,Gonzalez1989} method was used to assess that the localized TSs correctly connect to the corresponding minima along the imaginary mode of vibration. Enthalpies and Gibbs free energies were calculated at 298 K and 1 atm from total energies, including zero-point and thermal corrections, using standard statistical procedures \cite{McQuarrie1999}.

\section*{Results}

The first reaction that we used to test the method is
\begin{equation}
\label{eq:no}
2\mathrm{NO}+2\mathrm{H_2} \ \longrightarrow \ \mathrm{N_2} + 2\mathrm{H_2O}.
\end{equation}

Nitric oxide ($\mathrm{NO}$), a very stable compound having a bond dissociation energy of more than 600 kJ/mol, is one of the most relevant air pollutants that appear as a byproduct of combustion processes and has to be removed from effluent stream by reduction. To achieve this aim, a catalyst and a reducing agent are required, but a number of its known reaction channels tend to produce further pollutants. Today’s increased environmental awareness makes studying NO decomposition for its abatement particularly significant, with a special focus on the search for pollutant free mechanisms. 

Its reduction by $\mathrm{H_2}$ has been studied for decades \cite{Galvagno1978,Echigoya1980,Lee1986} and it has recently seen a surge of interest \cite{Bai2017,Pham2018} as the most desirable of its three reaction channels, leading to $\mathrm{N_2}$ and $\mathrm{H_2O}$ formation, has zero environmental impact \cite{Echigoya1980}. This reaction has been thoroughly investigated in presence of an assisting transition metal catalyst \cite{Bai2017,Pham2018}, whereas the reaction in the gas phase was examined only in a joint theoretical and experimental work \cite{Diau1995}. 
Theoretical modeling suggests the step $\mathrm{HNO + NO} \ \longrightarrow \ \mathrm{N_2O + OH}$ to be the rate-controlling one. It is also worth mentioning that, for calculating the reaction rate, the previous kinetic data by Hinshelwood and co-workers \cite{Hinshelwood1936} suggesting that the reaction is third-order with $-\mathrm{d} [\mathrm{NO}]/ \mathrm{d}t = k [\mathrm{H_2}] [\mathrm{NO}]^2$, was used.
Here we apply the HLDA procedure as described in the Methods section to gain a better understanding of the reaction mechanism. 

\begin{figure}[tb]
	\centering
	\includegraphics[width=1.0\columnwidth]{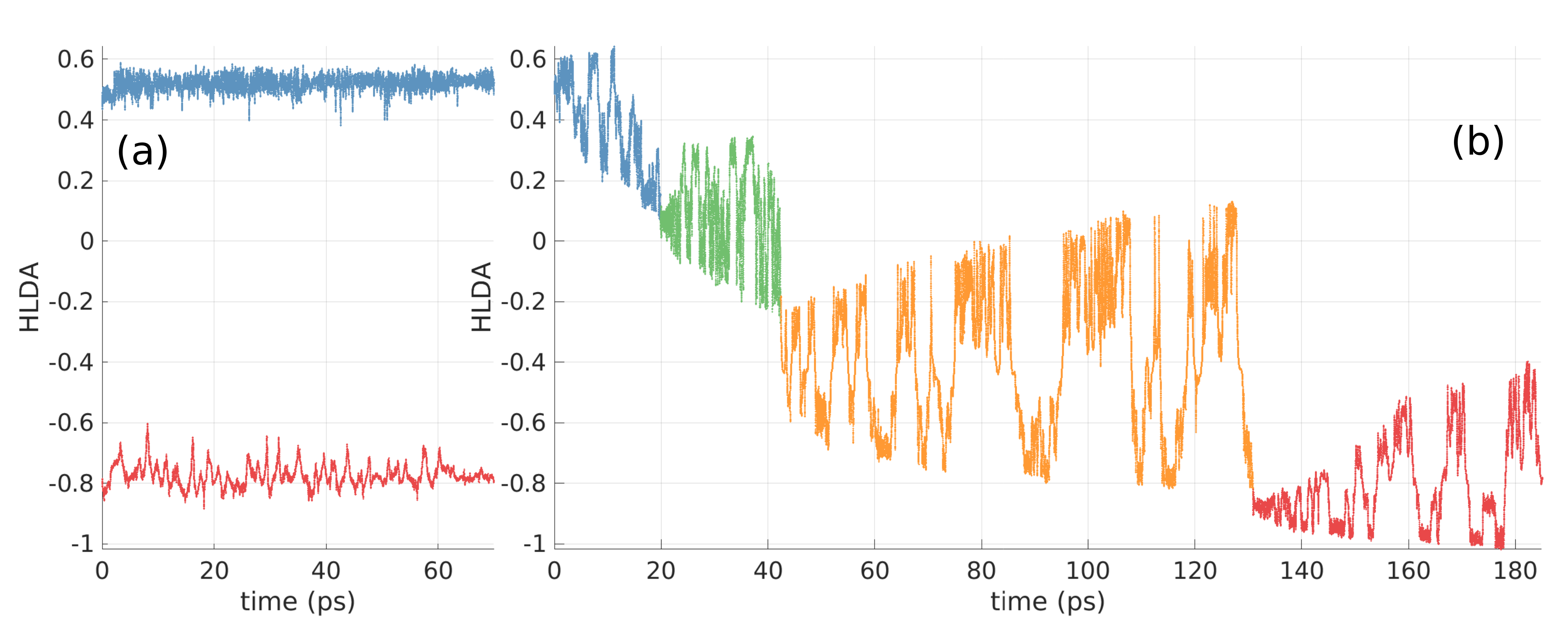}   
	\caption{In (a), dynamics of the HLDA variable for unbiased simulations of $\mathrm{2NO+2H_2}$ (in blue)  and $\mathrm{N_2+2H_2O}$ (in red). In (b), dynamics of the HLDA variable in a MetaD simulation starting from $\mathrm{2NO+2H_2}$ and, after crossing into two intermediates, reaching $\mathrm{N_2+2H_2O}$.
	}
	\label{fig:dynHLDA}
\end{figure}
In Fig. \ref{fig:dynHLDA} (a), the unbiased dynamics of the reactant state $2\mathrm{NO}+2\mathrm{H_2}$ (in blue) and the product state $\mathrm{N_2} + 2\mathrm{H_2O}$ (in red) is presented. The extent of the thermal fluctuations is limited and the HLDA variable is able to separate the two states effectively. In Fig. \ref{fig:dynHLDA} (b), we show the HLDA variable dynamics in the following MetaD simulation. In the SI we show the corresponding dynamics of the four most relevant descriptors by weight in the HLDA linear combination.

Starting from the reactants, as time grows, the bias deposition proceeds and the $\tilde{s}_{H}$ presents larger and larger fluctuations, which correspond to the nitric oxide molecules approaching each other and trying to form the metastable dimer compound $\mathrm{ONNO}$. 
The component needed for further stabilizing $\mathrm{ONNO}$ is the intervention of a hydrogen molecule that, by splitting and distributing one hydrogen atom at one end of $\mathrm{ONNO}$ and one on $\mathrm{N}$, leads to the formation of the intermediate state $\mathrm{HONNHO}+\mathrm{H_2}$ (in green). Early experimental work on a metal catalyst \cite{Gonzalez1970} suggested that the dimer of $\mathrm{NHO}$ is the most likely intermediate in the reduction of nitric oxide and a "likely precursor to the formation of $\mathrm{N_2O}$". Furthermore, $\mathrm{HONNHO}$ was identified as an intermediate in the aqueous reaction that leads to the formation of $\mathrm{N_2O}$.

Since the HLDA variable could distinguish the intermediate from the final product state, we let the MetaD simulation continue its exploration. In the next step, the hydrogen atom bound to $\mathrm{N}$ hops on the oxygen atom at the end of the chain triggering a cleavage of the corresponding $\mathrm{N-O}$ bond and the release of $\mathrm{H_2O}$. The resulting new intermediate state $\mathrm{N_2O}+\mathrm{H_2O}+\mathrm{H_2}$ (in yellow) is again distinguished from the final state and is also known experimentally to be an intermediate in the reaction in the presence of a metal catalyst \cite{Echigoya1980,Lee1986}. As the MetaD simulation proceeds further, we observe another reaction where $\mathrm{H_2}$ approaches the oxygen atom in $\mathrm{N_2O}$ and, by splitting, leads to the final product state $\mathrm{N_2} + 2\mathrm{H_2O}$. Remarkably, a complete pathway of the whole reaction, including its two intermediate states could thus be obtained in a single MetaD simulation. 

\begin{figure}[tb]
	\centering
	\includegraphics[width=0.72\columnwidth]{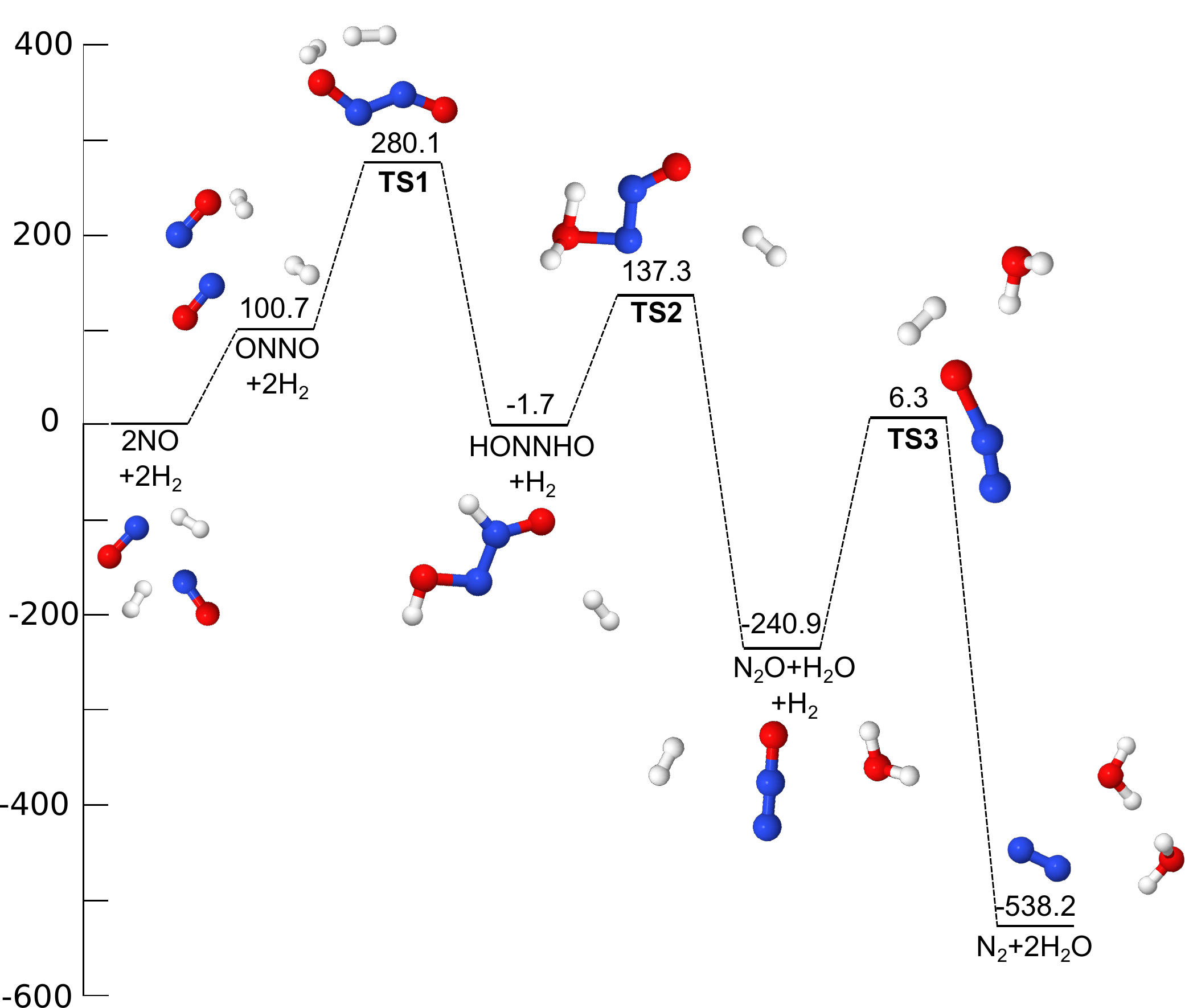}   
	\caption{Free energy profile (in kJ/mol) describing the $2\mathrm{NO}+2\mathrm{H}_2$ reaction obtained using DFT B3LYP static calculations. Along the reaction path, the structures of all intermediates and transition states are also sketched.
	}
	\label{fig:FES_ENergy_profile_NO_H2}
\end{figure}

Using the information from MetaD simulations, we run static quantum-mechanical calculations obtaining the free energy profile shown in Fig. \ref{fig:FES_ENergy_profile_NO_H2}.
We expect some inaccuracies in the quantitative description of the energetics of such reaction as a huge amount of theoretical investigations dealing with nitrogen oxides structures and thermodynamic quantities \cite{Sayos2000,Taguchi2008} demonstrated that DFT calculations give only approximate  estimates of such properties. 

Particularly, for the description of the structure and energetics of the $\mathrm{NO}$ dimer, an exemplary system of strong non-dynamic correlation, high-level \textit{ab-initio} computational methods should be used for providing more reliable results comparable with experimental data \cite{Glendening2007}. 
Using as zero reference energy the adduct formed by the $\mathrm{H_2}$ and  $\mathrm{NO}$ reacting species, the first step of the process is the endothermic, by 100.7 kJ/mol, formation  of the trans isomer of the  $\mathrm{ONNO}$ dimer preluding to the reaction with one of the  $\mathrm{H_2}$ molecules. 
The height of the barrier for the transition state TS1, corresponding to the splitting of the  $\mathrm{H_2}$ molecule and the formation of two new  $\mathrm{H-O}$ and  $\mathrm{H-N}$ bonds is 179.3 kJ/mol. The formed  $\mathrm{HONNHO}$ intermediate lies only 1.7 kJ/mol below the entrance channel. 

In the next step the $\mathrm{H}$ atom bound to nitrogen is transferred to the oxygen atom of the $\mathrm{OH}$ group to release water and form the $\mathrm{NNO}$ moiety. The energetic cost for the corresponding TS2 transition state is 139.0 kJ/mol, whereas the corresponding products together with the remaining $\mathrm{H_2}$ molecule lie 240.9 kJ/mol below the reactants’ energy.
The final step, that appears to be the rate-controlling step, leads to the formation of the final products, $\mathrm{N_2 + H_2O}$, surmounting an energy barrier of 247.2 kJ/mol. 
The negative frequency confirming the nature of first order saddle point of the intercepted structure corresponds to the breaking of the $\mathrm{H-H}$ bond and formation of two new $\mathrm{O-H}$ bonds, causing the detachment of a second water molecule. The whole reaction is calculated to be exothermic by 538.2 kJ/mol. It is worth underlining that such a reaction step sequence agrees with both the mechanism hypothesized in \cite{Gonzalez1970} and the third-order expression of the reaction rate reported in \cite{Hinshelwood1936}.


\begin{figure}[tb]
	\centering
	\includegraphics[width=0.92\columnwidth]{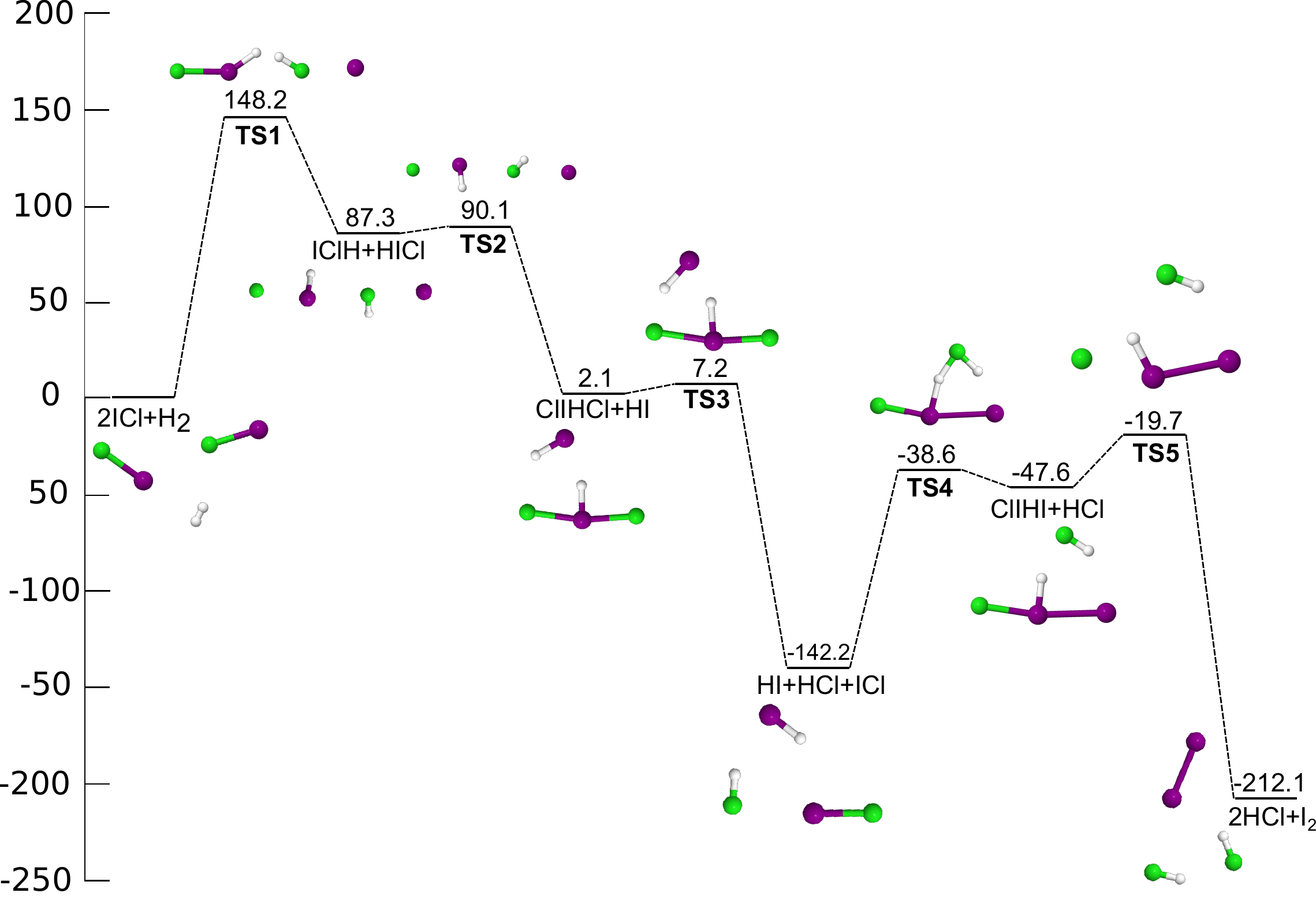}   
	\caption{Free energy profile (in kJ/mol) describing the $2\mathrm{ICl} + \mathrm{H_2}$ reaction obtained using DF B3LYP static calculations. Along the reaction path, the structures of all intermediates and transition states are also sketched.
	}
	\label{fig:FES_ENergy_profile_ICl}
\end{figure}

As a second example, we choose a paradigmatic reaction that appears often in chemistry textbooks
\begin{equation}
\mathrm{H_2} + 2\mathrm{ICl} \ \longrightarrow \ 2\mathrm{HCl} + \mathrm{I_2}
\label{eq:ICl}
\end{equation}
to demonstrate how reaction orders and reaction rate expressions cannot be inferred simply from the reaction equation. It is a common practice, therefore, to deduce the reaction mechanism from the rate law determined experimentally. We will show that, behind its apparent simplicity, reaction \ref{eq:ICl} hides a challenging and rich reaction mechanism which we are able to deduce without the need to use any experimental information.

Most works in the literature \cite{McDonald1972,Grosser1973,Loewenstein1985} date back to the 1970s-1980s and study its simpler analogous starting from $\mathrm{H} + \mathrm{ICl}$, commenting on the asymmetry in the prevalent formation of $\mathrm{HCl} + \mathrm{I}$ instead of $\mathrm{HI} + \mathrm{Cl}$. The most relevant difference with those studies is that in our case the reactant state includes a $\mathrm{H_2}$ molecule. Intuition says that the cleavage of the  $\mathrm{H-H}$ bond is necessary for the reaction to proceed, but it is not obvious how this should occur.

We performed the HLDA procedure and proceeded with MetaD simulations 
starting from  reactant and the product state. 
The dynamics of the HLDA variable during the unbiased and the MetaD simulations is presented in the SI.
We observed a reaction leading state $2\mathrm{HCl} + \mathrm{I_2}$ into the stable intermediate state $\mathrm{HI + HCl + ICl}$. 
There is no guarantee that the newly found states are well resolved by the HLDA CV used to discover them, as that variable has no prior information about them. In fact, in this case the intermediate state was not resolved effectively by the original HLDA variable. 
Therefore, continuing the MetaD simulations biasing the same variable would be fruitless, as a further deposition of bias would not contribute anymore for the exploration towards the final state. 

In this case one could either iterate the HLDA procedure determining a CV between the intermediate and the original final state or perform a multiclass HLDA. Here we chose the latter option (see SI for details). Multiclass HLDA includes information about all three states and separates them, generating 2 CVs. A MetaD simulation was performed biasing both CVs and their dynamics is presented in the SI.
The simulation revealed the mechanism connecting $\mathrm{H_2} + 2\mathrm{ICl}$ with the metastable state $\mathrm{ClIHI + HCl}$ and then $\mathrm{HI + HCl + ICl}$. Static calculations confirmed the mechanism and were used to create the reaction scheme in Fig. \ref{fig:FES_ENergy_profile_ICl}.

The whole reaction proceeds as follows.
Starting from the reactant state, the first step involves a cleavage of the $\mathrm{H_2}$ molecule assisted by the two $\mathrm{ICl}$ molecules that align to receive one $\mathrm{H}$ each, as is shown in \textbf{TS1} in Fig. \ref{fig:FES_ENergy_profile_ICl}. This leads to the formation of a metastable compound $\mathrm{ClIHCl}$ along with an intact $\mathrm{HI}$ molecule, which, after a few picoseconds, react further. We verified with unbiased simulations that $\mathrm{ClIHCl + HI}$ at our temperature is metastable and spontaneously decays into $\mathrm{HI + HCl + ICl}$. Successive static calculations confirmed that the barrier for this reaction is very low, about 5 kJ/mol.

The underlying reaction mechanism is quite complex as it involves the donation of an $\mathrm{H}$ atom from $\mathrm{HI}$ to $\mathrm{Cl}$ and the capture of another $\mathrm{H}$ from $\mathrm{ClIHCl}$. This triggers the cleavage of a $\mathrm{ICl}$ bond and the formation of the intermediate state $\mathrm{HI + HCl + ICl}$. We also intercepted an alternative TS involving only the $\mathrm{ClIHCl}$ molecule. It featured  the shift of the $\mathrm{H}$ atom from $\mathrm{I}$ to $\mathrm{Cl}$, which caused the cleavage of an $\mathrm{ICl}$ bond. 
 
Then, the reaction proceeds via a mechanism akin to the previous one. Through a double proton exchange, a metastable state $\mathrm{ClIHI + HCl}$ is formed. As before, unbiased simulations confirmed its spontaneous decay towards the final product state. Metadynamics simulations observed transitions into the final product state through two channels analogous to the previous ones. \textbf{TS5} in Fig. \ref{fig:FES_ENergy_profile_ICl} shows the bi-molecular one, as all the attempts to locate a TS for the tri-molecular mechanism via static calculations were unsuccessful.

The richness of the multiple atomic bonds rearrangements makes the investigated reaction a paradigmatic example of how the method is able to capture the complexity that hides behind seemingly trivial elements such as the nature of the initial and final states. Detailed insights are provided that can be translated into a precise sequence of steps.

\section*{Conclusions}

In this paper we present a method that is able to guess a chemical reaction pathway using a minimum amount of information about the reactant and the product states. We showed its capability by applying it to two reactions and finding the intermediates and the transition states involved in the mechanism. The method is general as it conducts its search for reaction pathways in free energy space, naturally taking into account entropic effects, and it can be applied to other kinds of multi-molecular reactions. 
We are currently working on its application to homogeneously and heterogeneously catalyzed reactions and reactions in a solvent. Investigations carried out adopting such a strategy might lead to the always desired control and fine tuning of chemical reactions by preliminary energy-saving and environmental-friendly $\textit{in silico}$ experiments.

\acknowledgement
We acknowledge the Swiss National Science Foundation Grant Nr. 200021\textunderscore169429/1 and the European Union Grant No. ERC-2014-AdG-670227/VARMET for funding.
Calculations were carried out on the ETH Z\"urich cluster Euler.

\begin{suppinfo}
	
The file \textit{Supporting\_info.pdf} contains further details about the simulations.
	
\end{suppinfo}

\newpage
\bibliography{library,PAPERS_USI_2017-HLDA,PAPERS_USI_2017-Metadinamica}

\end{document}


\maketitle
\date{\today}

\section*{Metadynamics simulations}

\noindent
In the first reaction
\begin{equation}
\label{eq:no}
2\mathrm{NO}+2\mathrm{H_2} \ \longrightarrow \ \mathrm{N_2} + 2\mathrm{H_2O}.
\end{equation}
we performed unbiased simulations for the reactant and the product state, respectively for $75$ ps and $100$ ps. The simulations were run in a cubic box with edge length $15 \ \text{\normalfont\AA}$ and periodic boundary conditions. To prevent the molecules from reaching the box edges and possibly interacting with unwanted periodic images, we confined them in a smaller region through harmonic restraints described by the potential $k (x - a)^2$ where $k = 200$ $\textrm{kJ\,mol}^{-1} \text{\normalfont\AA}^{-2}$, $a = 9 \text{\normalfont\AA}$ and x is the maximum distance between couples of atomic species. We applied the restraint between couples of nitrogen atoms, oxygen atoms and nitrogen and hydrogen atoms.

The HLDA procedure was run on the descriptor set based on the coordination number (CN) expression from Eq. 7 in the main text 
\begin{equation}
\label{Eq:Coordination_number}
q_{ij}(r_{ij})=\frac{1-\big(\frac{r_{ij}}{r_0}\big)^n}{1-\big(\frac{r_{ij}}{r_0}\big)^m}
\end{equation}
where $n=6$, $m=8$ and $r_0$ depends on the typical bond length between atoms $i$ and $j$. We tend to set it about $0.3 \ \text{\normalfont\AA}$ larger than the typical bond length to optimize the region of effectiveness of the switching function. For every couple of atomic species $A$ and $B$ in the system, the maximum $d_{\mathrm{max}}^{A-B}$ and minimum $d_{\mathrm{min}}^{A-B}$ of the CN is taken, according to Eq. 8-9 in the main text, except for the trivial case when one or two atoms belong to each species, as detailed in the main text. We point out that $d_{\mathrm{max(min)}}^{A-B} \neq d_{\mathrm{max(min)}}^{B-A}$, therefore both cases have to be included.

In this reaction, we used a set of $N_d = 14$ descriptors: ($d^{N-N}$, $d^{O-O}$,  $d_{\mathrm{max}}^{N-O}$, $d_{\mathrm{min}}^{N-O}$, $d_{\mathrm{max}}^{O-N}$, $d_{\mathrm{min}}^{O-N}$, $d_{\mathrm{max}}^{N-H}$, $d_{\mathrm{min}}^{N-H}$, $d_{\mathrm{max}}^{H-N}$, $d_{\mathrm{min}}^{H-N}$, $d_{\mathrm{max}}^{O-H}$, $d_{\mathrm{min}}^{O-H}$, $d_{\mathrm{max}}^{H-O}$, $d_{\mathrm{min}}^{H-O}$) with a corresponding set of $r_0$ = (1.5, 1.7, 1.6, 1.6, 1.6, 1.6, 1.4, 1.4, 1.4, 1.4, 1.4, 1.4, 1.4, 1.4) $\text{\normalfont\AA}$.

We applied the SVD decomposition to the covariance matrices by setting $N_f = 1$ singular value to zero and determined the HLDA linear combination of descriptors that generated the filtered HLDA CV $\tilde{s}_{H}$. The coefficients of the descriptors linear combination were (-0.39547, 0.06121, -0.20744, 0.60671, -0.11341, 0.47938, 0.02201, -0.03897, -0.00129, 0.00439, 0.03726, -0.13914, -0.08409, -0.39524). We then performed a metadynamics simulation starting from the reactant state by biasing $\tilde{s}_{H}$ with the Gaussian kernels from Eq. 1 with $\mathrm{W}=3$ kJ/mol, $\sigma=0.05$ kJ/mol and a deposition rate of one Gaussian every 100 timesteps.
The hill height values were chosen as usual with MetaD simulations to be in the ballpark of 1 kT. Given the computational cost of \textit{ab-intio} simulations and the exploratory aim of the method, to reduce the total cost of the simulations we choose higher than usual hill heights and deposition frequency to enable a quicker exploration of configuration space.

\begin{figure}[tb]
	\centering
	\includegraphics[width=1.0\columnwidth]{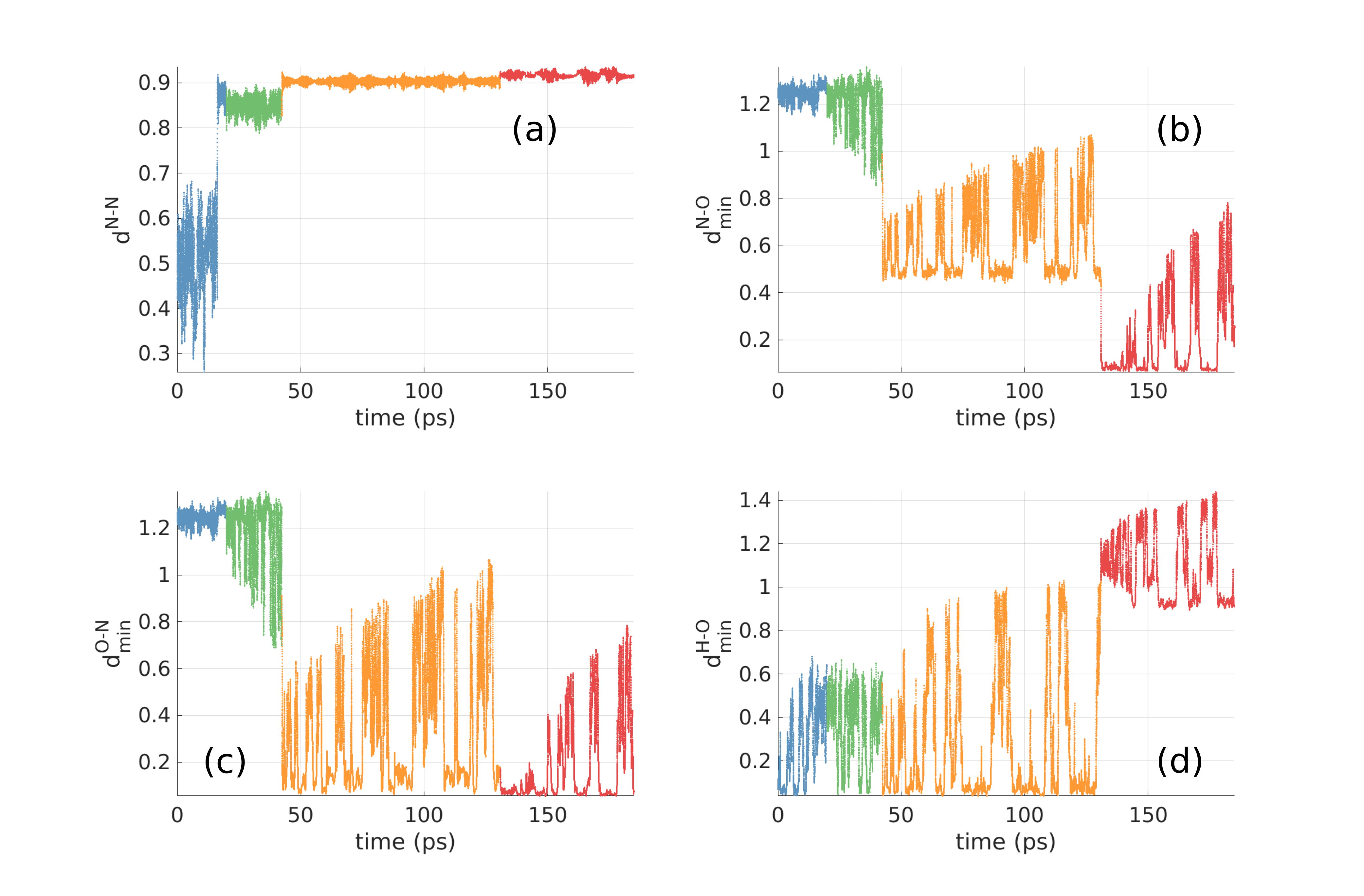}   
	\caption{The dynamics of the four most relevant non-normalized descriptors by weight in the HLDA linear combination during the MetaD simulation starting from $\mathrm{2NO+2H_2}$. In (a) there is $d^{N-N}$, in (b) $d_{\mathrm{min}}^{N-O}$, in (c) $d_{\mathrm{min}}^{O-N}$ and in (d) $d_{\mathrm{min}}^{H-O}$.
	}
	\label{fig:dyndescH2N2}
\end{figure}
The dynamics of the HLDA CV during the metadynamics simulation is shown in Fig. 2 (b) in the main text.
The corresponding dynamics of the most relevant descriptors is presented in Fig. \ref{fig:dyndescH2N2}. Descriptor $d^{N-N}$ is very effective at describing the first step of the reaction from $2\mathrm{NO}+2\mathrm{H_2}$ (in blue) to $\mathrm{HONNHO + H_2}$ (in green) as it captures the formation of a bond between the nitrogen atoms, $d_{\mathrm{min}}^{N-O}$ and $d_{\mathrm{min}}^{O-N}$ capture the following step where $\mathrm{N_2 O + H_2O + H_2}$ (in yellow) forms with the cleavage of a $\mathrm{N-O}$ bond, while $d_{\mathrm{min}}^{H-O}$ is able to describe the formation of a second water molecule during the last reaction step leading to $\mathrm{N_2 + 2 H_2O}$ (in red).

\noindent
In the second example, we considered reaction
\begin{equation}
\mathrm{H_2} + 2\mathrm{ICl} \ \longrightarrow \ 2\mathrm{HCl} + \mathrm{I_2}.
\label{eq:ICl}
\end{equation}
We ran unbiased simulations of the reactant and product state lasting respectively $54$ and $43$ ps. The simulations were run in the same cubic box as in the previous example and the same harmonic restraint was applied, this time acting on the maximum distance between couples of hydrogen and chlorine atoms, hydrogen and iodine atoms and couples of iodine and chlorine atoms. 

The descriptors set was made of $N_d = 15$ descriptors: ($d^{H-H}$, $d^{I-I}$, $d^{Cl-Cl}$, $d_{\mathrm{max}}^{H-Cl}$, $d_{\mathrm{min}}^{H-Cl}$, $d_{\mathrm{max}}^{Cl-H}$, $d_{\mathrm{min}}^{Cl-H}$, $d_{\mathrm{max}}^{H-I}$, $d_{\mathrm{min}}^{H-I}$, $d_{\mathrm{max}}^{I-H}$, $d_{\mathrm{min}}^{I-H}$, $d_{\mathrm{max}}^{I-Cl}$, $d_{\mathrm{min}}^{I-Cl}$, $d_{\mathrm{max}}^{Cl-I}$, $d_{\mathrm{min}}^{Cl-I}$) with a set of $r_0$ = (1.1, 3.0, 2.3, 1.6, 1.6, 1.6, 1.6, 1.9, 1.9, 1.9, 1.9, 2.6, 2.6, 2.6, 2.6) $\text{\normalfont\AA}$.

\begin{figure}[tb]
	\centering
	\includegraphics[width=1.0\columnwidth]{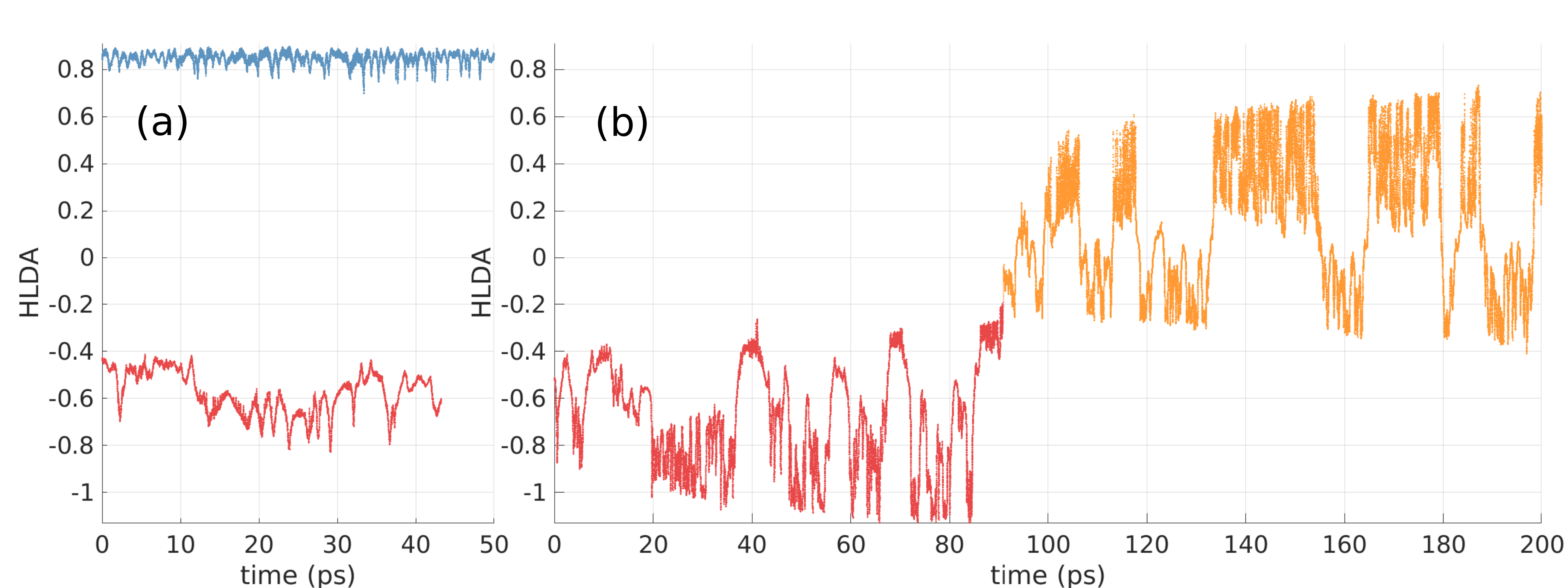}   
	\caption{In (a), dynamics of the HLDA CV for unbiased simulations of $\mathrm{H_2} + 2\mathrm{ICl}$ (in blue) and $2\mathrm{HCl} + \mathrm{I_2}$ (in red). In (b), dynamics of the HLDA variable in a MetaD simulation starting from $\mathrm{2\mathrm{HCl} + \mathrm{I_2}}$ which forms the intermediate state $\mathrm{HI + HCl + ICl}$ (in yellow).
	}
	\label{fig:dynICl1d}
\end{figure}

After the application of the SVD filtering with $N_f = 2$, the resulting linear combination of descriptors that generated the filtered HLDA CV was (0.7435, -0.0278, -0.3230, 0.0052, -0.4295, -0.0315, -0.3276, -0.0540, 0.0093, 0.0297, 0.0162, 0.1870, -0.0130, 0.0911, 0.0412). 
The dynamics of the HLDA CV for the reactant (in blue) and the product state (in blue) is presented in Fig. \ref{fig:dynICl1d} (a).

We set up exploratory metadynamics simulations starting from both the reactant and the product state, with $\mathrm{W}=1.5$ kJ/mol, $\sigma=0.05$ kJ/mol and a deposition rate of one Gaussian every 100 timesteps. In Fig. \ref{fig:dynICl1d} (b) it is shown the dynamics of the HLDA CV in a simulation starting from $2\mathrm{HCl} + \mathrm{I_2}$ (in red). The simulation discovered the intermediate state $\mathrm{HI + HCl + ICl}$ (in yellow), but it was stopped after reaching 200 ps simulation time as the HLDA CV almost reached the value of the $\mathrm{H_2} + 2\mathrm{ICl}$ unbiased state.

We performed a further unbiased simulation on the intermediate lasting $58$ ps with the aim of using it to improve the following simulations.
We then applied multiclass HLDA by including the newly found intermediate and determined two HLDA CVs: $\tilde{s}_{H1}$ with linear combination coefficients (-0.4446, 0.0049, 0.0288, 0.1219, 0.0686, 0.0886, 0.0323, 0.1243, 0.0676, -0.0530, -0.0781, 0.4969, 0.3439, -0.4511, -0.4189) and $\tilde{s}_{H2}$ with coefficients (-0.0450, 0.0427, 0.2969, -0.4559, 0.4762, -0.0061, 0.1473, -0.3869, -0.1632, 0.2848, 0.2102, -0.0487, 0.2557, -0.2622, -0.1160). 

\begin{figure}[tb]
	\centering
	\includegraphics[width=1.0\columnwidth]{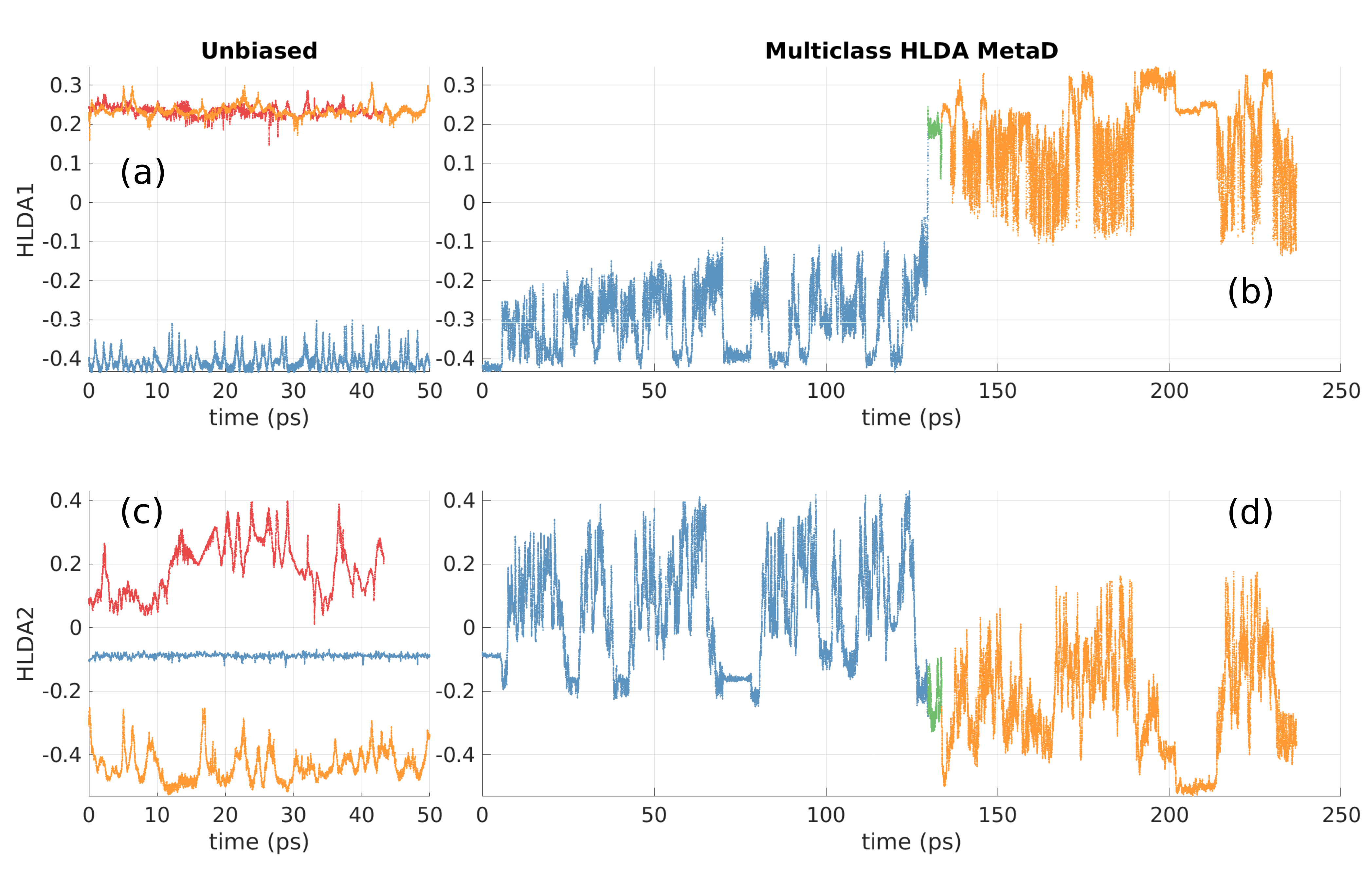}   
	\caption{In (a) and (c), dynamics of the two HLDA CVs during unbiased simulations of $\mathrm{H_2} + 2\mathrm{ICl}$ (in blue), $\mathrm{HI + HCl + ICl}$ (in yellow) and $2\mathrm{HCl} + \mathrm{I_2}$ (in red). In (b) and (d), the corresponding dynamics of the same HLDA variables during a MetaD simulation starting from $\mathrm{H_2} + 2\mathrm{ICl}$.
	}
	\label{fig:dynICl2d}
\end{figure}
The dynamics of $\tilde{s}_{H1}$ and $\tilde{s}_{H2}$ during the unbiased simulations is presented in Fig. \ref{fig:dynICl2d} (a) and (c) respectively.
Subsequent metadynamics simulations were run biasing both CVs with analogous deposition parameters as before. 
The dynamics of the two HLDA CVs during this MetaD simulation is shown in Fig. \ref{fig:dynICl2d} (b) and (d) respectively. We observed a reaction from  $\mathrm{H_2} + 2\mathrm{ICl}$ (in blue) into the metastable state $\mathrm{ClIHI + HCl}$ (in green) which further reacts into the known intermediate $\mathrm{HI + HCl + ICl}$ (in yellow), thus completing the mechanism scheme.
